\documentclass[pra,twocolumn,groupedaddress,showpacs]{revtex4}
\usepackage{graphics}
\usepackage{enumerate}
\begin{document}
\newcommand{\beq}{\begin{equation}}
\newcommand{\eeq}{\end{equation}}
\newcommand{\beqa}{\begin{eqnarray}}
\newcommand{\eeqa}{\end{eqnarray}}
\bibliographystyle{apsrev}

\title{Are weak measurements really necessary for Leggett-Garg type measurements?.}

\author{N.D. Hari Dass}
\email{dass@tifrh.res.in }
\affiliation{TIFR-TCIS, Hyderabad 500075, India.}


\begin{abstract}
Leggett-Garg inequalities are an important milestone in our quest to bridge the classical-quantum divide. An experimental investigation
of these inequalities requires the so called \emph{non-invasive measurements}(NIM). It has become popular to invoke weak measurements as
the means of realising NIMs to very good approximation, because of their allegedly low disturbance of systems under measurement. In this note, 
this is shown to be a myth; it is shown, by simple estimates of errors, that for comparable levels of statistical errors, even the strong or 
projective measurements can be used. In fact, it is shown that resource-wise, strong measurements are even preferable.

\end{abstract}

\maketitle

\section{Introduction.}
One of the most puzzling aspects of physics is a sharp dichotomy between the so called classical and quantum descriptions of 
physical systems. The classical description has proved itself extraordinarily succesful in the case of macroscopic objects(here
we use this term in an intuitive sense, without going into any precise characterization of what constitutes macroscopicity), while it
has proved to be utterly inadequate for the description of atomic and subatomic phenomena. Seemingly opposite appears to be the situation
with respect to quantum mechanics; while being extraordinarily succesful in the description of the microscopic world, the very
robustness of the macroscopic world is inscrutable from the quantum perspective. This dichotomy can be brought to sharper focus on
noting  that in the classical description i) systems have \emph{definite} properties, independent of any observations, and ii) observations,
equivalently measurements, do not alter the \emph{state} of the system, or do so in a way that one can compensate for the alterations to
restore the system to its original state; on the other hand, in quantum mechanics neither of the two is valid. The first becomes invalid
due to the occurrence of \emph{non-commuting observables}, and no state can have definite values for all of them. In fact, quantum theory,
at least the \emph{Copenhagen interpretation} even denies any meaning to a quantum state possessing definite values to any observable prior
to the process of measurement. Instead, it claims that the act of measurement produces outcomes for whose probabilities the theory makes
definite predictions. Furthermore, this interpretation of quantum theory asserts that generically measurements \emph{irretrievably} alter
the state of a system. It is a miracle of sorts that an ensemble of measurements, with sufficiently large ensembles, actually produces
enough information to know the initial state fully. Yet, if we view the macroscopic world as being built out of microscopic constituents,
this dichotomy is most disturbing. More specifically, in quantum theory a linear superposition, in a precisely defined sense, of physical
states is also a bona fide physical state, whereas in classical theory no sense can be attributed to such a superposition.

Leggett and Garg \cite{lgorig} sought to investigate if one could extend quantum theory to be `compatible' with classical theory. They intended such
extensions, called macrorealist by them, to be minimal. In particular, they required such extensions to satisfy(the precise terminology is
theirs): i) Macroscopic realism: effectively this requires macroscopic objects to be always in exactly one of the many states they can be in,
ii) Non-invasive measurements: these are measurements on macroscopic systems that can determine the (definite) state a macroscopic system is in
with arbitrarily small disturbance of the system. Additionally, they required a notion of \emph{induction}, which is somewhat technical and
not relevant in itself to the purposes of this article, which is to investigate how best such non-invasive measurements may actually be realised
in practical terms. As a test of macrorealism, they derived inequalities which outcomes of measurements must obey were the world
really macrorealist. These inequalities are in the same spirit as the well known Bell's inequalities for the so called realism. 
Some liken the LG-inequalities to temporal analogs of Bell's inequalities.
\section{Experimental setup}
The schema proposed by them, and their variants, can be found in \cite{lgorig,gargspl,home,homespl}. Some experiments that have been carried out 
in this context can be found in \cite{mahesh,maheshspl,palacios}. The experiment of Palacios is particularly relevant for our paper as it uses
weak measurements to perform these tests.
The idea is to perform measurements on series of subensembles of a larger ensemble of identical initial
states, prepared at, say, time $t_0$. In each of the series, some quantity, Q(t) is measured at two different times. In the macrorealist world, the system will be in the same
state at the later instant of measurement in each series as the measurement at the corresponding earlier instant is non-invasive. 
A moot point is whether an ensemble measurement on identically prepared
states is a reasonable thing to do, as in the macrorealist world all assertions pertain to individual systems. There are also subtle issues
relating to the homogeneity of the ensembles as perceived in the macrorealist world even though they all represent homogeneous ensembles
in the quantum world \cite{neumannbook}.

It is evident that some form of quantum measurement that best approximates a non-invasive measurement, whatever that means, is needed in 
any experimental
setup seeking to test the Leggett-Garg concept of macrorealism. Assuming that it is possible to do so, the schema goes on to specify
that in ${\mathrm k}$ series, measurements should be performed at times $t_1 < t_2 < t_3 < \ldots < t_{\mathrm k}$ such that in the first series they are performed
at $(t_1,t_2)$, in the second at $(t_2,t_3)$ etc., and finally, in the ${\mathrm k}$-th at $(t_1,t_{\mathrm k})$. The two independent
measurements at $t_1$
have to be non-invasive, but neither of the two independent measurements at $t_{\mathrm k}$ need be non-invasive . At each of the other 
instants of time $(t_2,\ldots,t_{\mathrm k-1})$, only the earlier of the two independent measurements have to be necessarily 
non-invasive. That is, measurement at $t_2$ of the second of the series, and, at $t_3$ of the third of the series etc., have to be non-invasive.
Thus exactly half the total of ${\mathrm 2k}$ measurements need be non-invasive.
The exact form of these inequalities is irrelevant to our considerations. Our focus is entirely on the issue of what will be good candidates
for the non-invasive measurements required by the schema. Before proceeding, we give a mathematical criterion for invasiveness which ought
to be reliable in generic circumstances.
\subsection{A criterion for invasiveness}
It is intuitively obvious that an invasive measurement alters in some strong sense the state of the system that is being measured. But needless to say, it is useful to have a quantitative measure of such invasiveness. Let us consider the case where the system, before measurement, is
described by the density matrix of a pure state $\rho_{ini}$. Being pure, it satisfies $\rho_{in}^2 = \rho_{in}$. In particular, $tr\,\rho_{in}^2=1$. The \emph{purity} ${\cal P}$ of the state is defined as ${\cal P} = tr \rho^2$. Thus the initial purity is ${\cal P}_{ini} =1$. The post-measurement density matrix is \emph{mixed} in the sense that $\rho_{post}^2 \ne \rho_{post}$. The purity of mixed states is always less than 
unity i.e ${\cal P}_{post} \le 1$. The change in purity due to measurement is one measure of the invasiveness denoted here by ${\cal I}$.
Therefore ${\cal I}_1 = {\cal P}_{ini}-{\cal P}_{post}$. Here we have introduced a label to indicate that this is one of the many possible
ways of quantifying invasiveness. 

Yet another measure is the the departure from unity of the overlap of the post-measurement state with 
the initial state, often called the
\emph{fidelity}; denoting it by ${\cal F}$, it is given by ${\cal F} = tr\:\rho_{ini}\cdot\rho_{post}$. As we shall see later on, purity and
fidelity can sometimes equal each other, but not always. Hence a second measure of invasiveness is given by 
${\cal I}_2 = 1 - tr\:\rho_{ini}\cdot\rho_{post}$.

Another aspect we will be concerned with is a measure of how much resource gets wasted as a result of the original ensemble being
'corrupted' by invasive measurements. For example, in the well known projective measurements, the state changes dramatically and the
entire ensemble used in measurement can be considered a resource wasted for the future. Therefore, we shall adopt the attitude that where
invasiveness measures ${\cal I}_i$ are of order unity, the entire ensemble used in the measurement is a wasted resource. If, on the other
hand, the invasiveness measures are much smaller than unity, we shall take it that a fraction ${\cal I}_i$ of the ensemble size used in the
measurement is to be treated as a wasted resource. 
\section{Brief resume of strong and weak measurements in quantum theory.}
As there is voluminous literature on both these types of measurements, we just summarise their features central to our considerations of
this article. For the sake of simplicity, we take the initial state of the system to be described by the pure density matrix $\rho_{ini}$. 
It is straightforward to generalize these considerations to cases where the initial state of the system is described by a mixed density matrix
\cite{nori}. In the so called strong measurements, also called projective measurements, measurement of an observable, say, ${\cal A}$,
with the spectrum $(a_i,|a_i\rangle)$ (assumed non-degenerate, for simplicity), in
an initial state described by the density matrix $\rho_{ini}$ of the system results in the post-measurement mixed state $\rho_{post}^{strong}$
given by
\begin{eqnarray}
\label{eq:strongpost}
\rho_{post}^{strong} &=& \sum_i\:tr\rho_{ini}\Pi_i\:\Pi_i\quad\quad \Pi_i = |a_i\rangle\langle a_i|\nonumber\\
&=& \sum_i\: {\tilde p}_i\,\Pi_i\quad\quad {\tilde p}_i = tr\:\rho_{ini}\,\Pi_i
\end{eqnarray}
The purity of the most-measurement state (and hence ${\cal I}_1$) is easily calculated to be:
\begin{equation}
\label{eq:strongpurity}
{\cal P}^{strong} = \sum_i\: {\tilde p}_i^2 \rightarrow {\cal I}_1^{strong} = 1-\sum_i\:{\tilde p}_i^2
\end{equation}
The fidelity ${\cal F}$ (and hence ${\cal I}_2$) is also easy to calculate:
\begin{equation}
\label{eq:strongfidelity}
{\cal F}^{strong} = \sum_i\:{\tilde p}_i^2 = {\cal P}^{strong}\rightarrow\:{\cal I}_2^{strong} = {\cal I}_1^{strong}
\end{equation}
Hence in the case of strong measurements, both measures of invasiveness agree. Since these are generically of order unity, the entire 
(sub)ensembles used for strong measurements are wasted resources.

Operationally, in strong measurements, the initial state of the apparatus is the so called pointer state $|P_0\rangle_A$, which is a state
with the pointer variable ${\mathrm p}$ having, say, zero mean, and very small dispersion(variance). 
In contrast, in a so called weak
measurement, the initial state of the apparatus is taken to be one having, say, zero mean for the pointer variable, but having a very 
large variance $\Delta_p$. Consequently, it is not a pointer state. But with carefully defined pointer states, it can be taken to be a 
superposition of a very large number of pointer states $|P_i\rangle_A$ with ${\mathrm p}=i$ as the mean value of the pointer variable and 
again with very small variance. The measurement interaction can be chosen to be the same for both types of measurements. For example, for
qubit systems, a typical measurement interaction is one which would shift the pointer state $|P_0\rangle$ to $|P_1\rangle$ if the system
was in the state $|\uparrow\rangle_S$, and to $|P_{-1}\rangle$ if the qubit was in state $|\downarrow\rangle_S$.
The post-measurement density matrix of the system is given by \cite{weaksingle}:
\begin{equation}
\label{eq:weakpost}
\rho_{post}^{weak} = \rho_{ini} - \frac{1}{4\Delta_p^2}\:\sum_{i,j}\,\alpha_i\,\alpha_j^*\,(a_i-a_j)^2\,|a_i\rangle\langle a_j|+\ldots
\end{equation}
The dots in this equation denote terms higher order in $\Delta_p^{-2}$. Here, the initial pure state density matrix, expressed in the basis $|a_i\rangle$ is taken to be:
\begin{equation}
\label{eq:initialstate}
\rho_{ini} = \sum_{i,j}\:\alpha_i\,\alpha_j^*\:|a_i\rangle\langle a_j|
\end{equation}
It is straightforward to work out the purity of the post-measurement state in the case of weak measurements:
\begin{equation}
\label{eq:weakpurity}
{\cal P}^{weak} = 1 - \frac{2}{4\Delta_p^2}\:\sum_{i,j}\:{\tilde p}_i\,{\tilde p}_j\,(a_i-a_j)^2+\ldots 
\end{equation}
Therefore, the invasiveness measure ${\cal I}_1$ for weak measurements turns out to be
\begin{equation}
\label{eq:weakinvI}
{\cal I}_1^{weak} = 
\frac{2}{4\Delta_p^2}\:\sum_{i,j}\:{\tilde p}_i\,{\tilde p}_j\,(a_i-a_j)^2+\ldots 
\end{equation}
Likewise, the fidelity for weak measurements is:
\begin{equation}
\label{eq:weakfidelity}
{\cal F}^{weak} = 
{\cal P}^{weak} = 1 - \frac{1}{4\Delta_p^2}\:\sum_{i,j}\:{\tilde p}_i\,{\tilde p}_j\,(a_i-a_j)^2+\ldots 
\end{equation}
Finally, the second invasiveness measure for weak measurements follows from the above equation, and is given by:
\begin{equation}
\label{eq:weakinvII}
{\cal I}_2^{weak} = \frac{2}{4\Delta_p^2}\:\sum_{i,j}\:{\tilde p}_i\,{\tilde p}_j\,(a_i-a_j)^2+\ldots\simeq 2\,{\cal I}_1^{weak} 
\end{equation}
One sees that for weak measurements ${\cal I}_1^{weak} = 2\,{\cal I}_2^{weak}$, while for strong measurements these were the same.
\section{Application to LG-measurements}
It is instructive to recall both the average outcomes as well as the errors in the two types of measurements. In strong measurements, 
every outcome of the pointer position, after suitable shifting and scaling, is one of the eigenvalues $a_i$;each occurring with 
probability ${\tilde p}_i$ so that the average of the pointer positions is
\begin{equation}
\label{eq:strongmean}
\langle\,{\mathrm p}\, \rangle_{strong} = 
\sum_i\,{\tilde p}_i\,a_i = tr\:\rho_{ini}\,A=\langle\:A\:\rangle_{ini} 
\end{equation}
In strong measurements, the state of the system, when the outcome of the pointer position corresponds to the eigenvalue $a_i$ of
the observable, is $\Pi_i=|a_i\rangle\langle a_i|$. In such a state, the observable A has the definite \emph{value} $a_i$. It is
in this sense that one can talk about the outcomes of strong measurements being the eigenvalues of the measured observable. The
variance of the statistical distribution of outcomes in strong measurements is given by:
\begin{equation}
\label{eq:strongvar}
(\Delta p)^2_{strong}= \sum_i\,{\tilde p}_i\,a_i^2 - \langle\:A\:\rangle_{ini}^2 = (\Delta\:A)^2_{ini}
\end{equation}
In weak measurements, the situation is dramatically different. The outcomes of the pointer positions are many and generically distinct from the
eigenvalues. In fact, the outcomes sample the entire range of pointer positions, which, by virtue of the nature of the initial apparatus state,
is very large. Equally important, the state of the system associated with a particular pointer variable outcome, is generically not an eigenstate of the observable at all. Consequently, there is no value of the observable that can even remotely be associated with the outcomes of 
weak measurements. Rather remarkably, the average of the pointer positions, after the same shifting and rescaling, turns out to be the
same as in strong measurements!
\begin{equation}
\label{eq:weakmean}
\langle\,p\,\rangle_{weak} = \langle\:A\:\rangle_{ini}
\end{equation}
This can lead to the (false)hope that complete information about an arbitrary quantum system may be obtained with arbitrarily small
disturbance of it, in the sense of arbitrarily small invasiveness, in the precise sense outlined above. What dashes this hope is the 
variance in the observed distribution of pointer positions \cite{weaksingle,nori}:
\begin{equation}
\label{eq:weakvar}
(\Delta p)^2_{weak} = \frac{\Delta_p^2}{2} + (\Delta\:A)^2_{ini}
\end{equation}
This equation tells that the outcomes in weak measurements have no bearing on the values of the observables. Not only is this connection
absent on an event to event basis, it is strongly absent in the statistics of the outcomes too (except for the mean). Since the information 
about the system is
provided by the mean of the distribution, the extreme largeness of $\Delta_p$ makes weak measurements very very noisy, and large enough 
ensemble measurements are needed to bring down the statistical errors sensibly. While this obvious fact has been noted in the current 
literature, exactly how large the ensembles ought to be, and what other effects such large ensemble measurements end up producing, are 
not precisely stated. In the context of LG-measurements, these concerns are what have been addressed in this note.

Now we compare the relative efficacies of strong and weak measurements in so far as LG-measurements are concerned. It makes no sense
to compare experimental setups yielding different accuracies of measurements. Hence we stipulate the errors in all measurements to
be $\epsilon$ before hand. We shall adopt the statistical errors to be the criteria for our comparisons (a first account of these ideas
was mooted in \cite{ndhspl}). This could well have its
limitations, but it is one of the simplest of criteria. Let the size of the initial ensemble of identically prepared states be M. In
the version with k time slices \cite{lgorig}, we divide this into k subensembles of size $\frac{M}{k}$ each. So far, this is just following
the original schema. According to that, on each such subensemble one weak measurement(the earlier) and one strong measurement are to
be performed. One could have opted for performing weak measurements both the times. Then the available sub-subensemble size for each weak
measurement would have been $\frac{M}{2k}$, and the resulting error in each weak measurement would have been ${\tilde\epsilon}_w
= \frac{\Delta_p}{\sqrt{2\frac{M}{2k}}}=\frac{\Delta_p}{\sqrt{\frac{M}{k}}}$. For the same ensemble size, errors in strong measurements
are down by a factor of $\frac{\sqrt{2}(\Delta\,A)_{ini}}{\Delta_p}$, which is really negligible if the weak measurements are truly weak.
Else, they can not be taken to be non-invasive even in the sense described here. Therefore, it makes sense to make one of the measurements 
strong, so a larger sub-subensemble can be made available to the weak measurement, thereby reducing the statistical error in it. This way,
the error in the weak measurement becomes $\sqrt{2}$ times smaller and we can set $\epsilon = \frac{\Delta_p}{\sqrt{\frac{2M}{k}}}$.

Now we propose our schema which is to do all the measurements with only projective measurements. We first analyse the sub-subensemble size 
required to 
deliver, for each strong measurement, the same error $\epsilon$ given in the previous para. That is given by
\begin{equation}
\label{eq:strongsubensemble}
M_s = \frac{(\Delta\,A)^2_{ini}}{\epsilon^2} = \frac{(\Delta\,A)^2_{ini}}{\Delta_p^2}\cdot\frac{2M}{k}
\end{equation}
As now a total of 2k strong measurements need to be performed, the total ensemble size required is
\begin{equation}
\label{eq:strongensemble}
M_{tot} = 2k\cdot \frac{(\Delta\,A)^2_{ini}}{\Delta_p^2}\cdot\frac{2M}{k} = \frac{4(\Delta A)^2_{ini}}{\Delta_p^2}\:M
\end{equation}
Thus the ensemble size required for checking LG-inequalities with strong measurements is much much smaller than what is required had
one used weak measurements! Another way of stating this is that had we used the same size of the ensemble(M) for all-strong measurements,
the errors in the LG-inequalities could have been reduced by enormous factors of $\frac{\sqrt{2}\,(\Delta\,A)_{ini}}{\Delta_p}$!
\subsection{Resource wastages}
Now we address the issue of resource wastages in the two schemes of testing LG-inequalities. It may appear at first that our schema of
using only strong measurements will lead to heavy loss of resources as, according to our criteria, strong measurements being invasive would 
render the entire ensemble used for them a waste. In contrast, weak measurements whose invasiveness can be lowered arbitrarily, would hardly
lead to any wastages. The flaw in this superficial appearance lies in having overlooked the role of errors. As argued till now, for comparable
errors, weak measurements require substantially larger ensembles. So even a tiny fraction of these huge ensembles lead to considerable
wastages. We now make this more quantitative.

For this, let us consider where we have a total ensemble of M which has been divided into k subensembles of size $\frac{M}{k}$. In the
original schema, on each of these one weak  measurement and one strong measurement are to be performed. But according to our proposal
on each of them two strong measurements are to be performed. But the errors in all measurements are to be kept at the same level i.e $\epsilon$.
The part of the subensemble to be used for weak measurement is, as we saw earlier, almost the entire subensemble of $\frac{M}{k}$. 
In the case of weak measurements, a fraction ${\cal I}^{weak}$ of the subensemble size gets lost as wasted resource. Using the more favorable 
invasiveness index ${\cal I}_1^{weak}\simeq\,\frac{1}{4\Delta_p^2}$, leads to a resource loss per weak measurement of  
$\simeq\,\frac{M}{k}\cdot\frac{1}{4\Delta_p^2}$. For the same level of statistical error, only an sub-ensemble size of 
$\frac{2(\Delta\,A)^2}{\Delta_p^2}\cdot\frac{M}{k}$ is needed per strong measurement.In the worst case, all of this is to be considered a
wasted resource. If the observable is qubit-like, $(\Delta\,A)^2\,\le\,\frac{1}{4}$. Thus the wasted resources are actually comparable. Though
the fraction of the ensemble required to be discarded in weak measurements is very very small, this is almost exactly compensated by the
huge sizes of ensembles required for weak measurements of comparable accuracies.  
\section{Discussion}
We have shown that when errors are properly taken into account, weak measurements have no advantages over strong measurements when it
comes to experimental investigation of Leggett-Garg inequalities. In fact, weak measurements fare far worse because of the humongous
resources called for. We have used ensemble size as an obvious resource. We have not even counted other measurement resources. Furthermore,
the wastage of resources as a result of the corruption of the state due to measurements are actually comparable in both cases. Once again, the 
low invasiveness of weak measurements gets compensated by the much larger ensemble sizes required for it. In a certain sense, substituting
strong measurements seems to have another benefit. As long as the quantum mechanical pure ensembles are homogeneous(a very basic assumption in
quantum theory \cite{neumannbook}), the subensembles for strong measurements are strictly identical, whereas in the version using weak measurements, the
subensemble available for the subsequent strong measurement in every series is actually mildly different from the subensemble used for
the corresponding weak measurements. Ironically, strong measurements mimic non-invasiveness even better! It is very important to redo
experiments on LG-inequalities in the light of our remarks and experimentally vindicate our stand. It must be mentioned that we have
only looked at weak measurements as possibilities for non-invasive measurements; we have not considered other variants based on the use of
ancillae as in \cite{mahesh}, or, negative result measurements (NRM) as in \cite{home}. We need to have a better understanding of the
errors in such measurements.
\acknowledgments{ I thank T.S. Mahesh and D. Home for many discussions. An illuminating expose of macrorealism by Anupam Garg\cite{gargspl}
was very helpful. Support for this work also came from DST(India) project IR/S2/PU-801/2008. }

\end{document}